# Spin Relaxation in Ge/Si Core-Shell Nanowire Qubits

Yongjie Hu[1,2†], Ferdinand Kuemmeth[2], Charles M. Lieber[1,3], Charles M. Marcus[2]

[1]*Department of Chemistry and Chemical Biology,* [2]*Department of Physics,* [3]*School of Engineering and Applied Sciences, Harvard University, Cambridge, Massachusetts 02138, USA.* [†]*Present Address: Department of Mechanical Engineering, Massachusetts Institute of Technology, Cambridge, Massachusetts, 02139, USA.*

**Controlling decoherence is the most challenging task in realizing quantum information hardware[1-3]. Single electron spins in gallium arsenide are a leading candidate among solid-state implementations, however strong coupling to nuclear spins in the substrate hinders this approach[4-6]. To realize spin qubits in a nuclear-spin-free system, intensive studies based on group-IV semiconductor are being pursued. In this case, the challenge is primarily control of materials and interfaces, and device nanofabrication. We report important steps toward implementing spin qubits in a predominantly nuclear-spin-free system by demonstrating state preparation, pulsed gate control, and charge-sensing spin readout of confined hole spins in a one-dimensional Ge/Si nanowire. With fast gating, we measure $T_1$ spin relaxation times in coupled quantum dots approaching 1 ms, increasing with lower magnetic field, consistent with a spin-orbit mechanism that is usually masked by hyperfine contributions.**

Since Loss and DiVincenzo's proposal[1], the promise of quantum dots for solid state quantum computation has been underscored by the successful initialization, manipulation and readout of electron spins in GaAs systems[5,7-9]. The electronic wave functions in these systems typically overlap with a large number of nuclear spins that are difficult to control and in most cases thermally randomized. The resulting intrinsic spin decoherence rates[4-6] have been successfully reduced by spin-echo techniques[6,10] but require complex gate sequences that complicate multi-qubit operations[11]. The prospect of achieving long coherence times in group IV materials with few nuclear spins has stimulated many proposals[12,13] and intensive experimental efforts[14-18], and crucially depends on the development of high-quality host materials. Recent advances on single quantum dots have been achieved using Zeeman splitting for readout at finite magnetic field[18-22]. Coupled quantum dot devices[12,14-17,23] in a nuclear spin-free system are more desirable for flexible quantum manipulation[24] but more challenging, and the characterization of the spin lifetime is still missing.

Semiconductor nanowires (NWs) are a favourable platform for quantum devices due to precise control of diameter, composition, morphology and electronic properties during synthesis[25]. The prototypical Ge/Si nanowire heterostructure has revealed diverse phenomena at the nanoscale and enabled numerous applications in nanoelectronics. The epitaxial growth of Si around a single-crystal Ge core (inset, Fig. 1) and the associated valence band offset provide a natural radial confinement of holes that – due to the large subband spacing – behave one-dimensional (1D) at low temperature[26]. Although the topmost valence band has been predicted to be two-fold degenerate and of light-hole character under idealized approximations[27], it is generally expected that mixing between heavy and light hole bands due to confinement will affect spin relaxation via phonons and spin-orbit interaction. To assess the potential of holes in Ge/Si heterostructure nanowires for spintronic applications we directly probed the spin relaxation times in top-gate defined quantum dots.

Our spin qubit device (Fig. 1) consists of a double quantum dot and charge sensor that is electrically insulated but capacitively coupled to the double dot via a floating gate (green). The double dot is defined by barriers in the nanowire induced by positive voltages applied to gates L, M, and R (purple). Plunger gates LP and RP (orange) tune the energy levels of each dot and gates EL and ER (magenta) increase carrier densities near source ($S_{dd}$) and drain ($D_{dd}$) in order to provide efficient charge transmission between nanowire and contacts. A tuning gate S (yellow) operates the sensor dot near a Coulomb oscillation where the sensor conductance $g_s$ (measured by ~ 30 $\mu V_{rms}$ excitation standard locking techniques) is most sensitive to the potential of the floating gate.

We first perform transport spectroscopy in a single quantum dot defined by top gates L and M, while leaving all other gate electrodes at smaller gate voltages to allow hole flow through the nanowire. Differential conductance $g_{dd}=dI/dV_{SD}$ as a function of bias $V_{SD}$ and gate voltage $V_{LP}$ (Fig. 2a) reveal Coulomb diamonds and the excitation spectrum of the dot. The data exhibits a set of conductance lines parallel to the diamond edge, due to tunnelling through asymmetric barriers into excited quantum states (inset, Fig. 2c). First, we note that for $V_{SD} > 0$ the lowest excited state appears about 0.6 meV above the ground state resonance. Such a high level spacing constitutes an advantage for spin qubits where the lowest spin states need to be addressed without occupation of higher orbitals. Second, a finite magnetic field splits the spin-degenerate ground state transition into a spin-up and spin-down resonance (Fig. 2b). Their separation allows us to extract the Zeeman energy $E_Z = |g| \mu_B B$, where $g$ is the g-factor and $\mu_B = 5.8 \times 10^{-5}$ eV/T is the Bohr magneton. A linear fit to Zeeman splittings at different magnetic fields (Fig. 2c, $g \sim$ 1.02) yields a g-factor that is significantly smaller than that of unperturbed light holes in germanium ($|g_{||}| \sim 2|\kappa| \sim 2 \times 3.41 = 6.82$)[27,28], likely due to strong confinement and heavy-hole light-hole mixing. While $g \neq 2$ provides indirect evidence for the presence of spin-orbit coupling, the observed two-fold spin degeneracy is consistent with recent experiments on holes in Ge and Si quantum dots[18,19,29,30]. Note that in the absence of external or internal (i.e. nuclear) magnetic fields hole states with half integer total angular momentum are expected to be (at least) two-fold degenerate due to time reversal symmetry, and hence it is natural to create spin qubits from such Kramers doublets in close analogy to the singlet-triplet qubits of two-electron spin states.

To form a double quantum dot in the spin-blockade regime[24] we raise the barriers and lower the hole density by tuning gate voltages. We estimate the number of holes in each dot to be between 10 and 50. In this regime, we verify the even-odd filling of spin-degenerate ground states using the charge sensor described previously[15]. Figure 3a shows the sensing data of six subsequent charge transitions of the double dot, visible as discrete peaks in the differential sensor dot conductance $dg_S/dV_L$. With increasing magnetic field the transitions increase or decrease in energy (gate voltage), depending on the direction of the added hole spin. To clarify this field dependence we plot the voltage spacing $\Delta V$ between consecutive peaks, reduced by their spacing near zero field (Fig. 3b). Assuming that the coupling efficiency of gate L is close to that extracted from Fig. 2a, we find that the ground state magnetic moments are consistent with $g$ = 1.0 (solid line, Fig. 3c). The alternating sign of magnetic moments reflects the even-odd filling of spin-degenerate hole states in the double dot. To simplify the following discussion we denote odd occupation of both dots as (1,1) and even occupation as (0,2), similar to singlet-triplet qubits in few-electron double dots[24]. Moreover, for sufficiently large orbital level spacing, the ground state of (0,2) is singly degenerate (all spins in each dot are paired, "singlet"), whereas (1,1) additionally allows parallel alignment of the two unpaired spins ("triplets").

Next, we characterize hole spin relaxation by realizing spin-to-charge conversion in a cyclic gate-pulse sequence E→R→M→E (see Fig. 4b inset) that is continuously repeated while measuring $g_S$ as a function of gate voltages $V_L$ and $V_R$. 80% of the cycle period is spent at the measurement point (M), so that the time-averaged sensor conductance reflects the charge state of the double dot during the measurement time $\tau_M$. Starting with the left dot empty, (0, 1), the pulse sequence resets the (1, 1) double dot during R to either a triplet or singlet state by tunneling a hole of random spin state through the left barrier. An adiabatic gate pulse to M converts only the (1, 1) singlet state to the (0, 2) charge configuration. States with parallel hole spins remain in the (1, 1) charge configuration due to Pauli exclusion, thereby increasing the sensor conductance $g_S$ within the metastable region outlined by a white dotted triangle. Within this pulse triangle, $g_S$ takes values between (1, 1) and (0, 2), characterized by the visibility $I = [g_S(\tau_M) - g_S(\infty)]/[g_S(0) - g_S(\infty)]$. With increasing measurement time, spin relaxation processes into the (0, 2) ground state are more likely to occur, resulting in a decay of the visibility with increasing $\tau_M$ (Fig. 4a). By fitting the measured visibility in the center of the pulse triangle by[5] $I(\tau_M) = (\tau_M)^{-1} \int_0^{\tau_M} e^{-t/T_1} dt$, we extract a spin relaxation time of $T_1 \sim 0.6$ ms at $B$ = 0 (Fig. 4b). At all measurement points, we observed no pulse triangle when the cycle is reversed (R→E→M→R), demonstrating that the asymmetric inter-dot tunnelling was dominated by Pauli blockade.

The extracted relaxation time $T_1$ decreases with increasing magnetic field to 0.3 ms and 0.2 ms at $B = 0.1$ and 1 T respectively (Fig. 4b). This weak dependence is qualitatively consistent with spin relaxation via spin-orbit interaction and two-phonon processes[31] but may also reflect relaxation through Coulomb coupling with electrons in the gates[32]. Additional work remains to illuminate the relaxation mechanism in detail, but we note that recent experiments in silicon and Si/SiO$_2$ interfaces have observed a clear power law dependence only at high magnetic fields (several tesla)[20,22,33]. Based on the split-off band in bulk Ge (0.29eV compared to 0.38eV in InAs[34]) we speculate that spin-orbit coupling is of comparable strength[13,35] as in InAs nanowire qubits[36] and, in conjunction with heavy-light hole mixing, may allow rapid electrical spin manipulation techniques that are absent in electron-based qubits, such as rapid direct spin rotation[37] and a recently predicted direct Rashba spin orbit interaction[13].

In summary, we have observed spin doublets of holes confined in one-dimensional Ge/Si nanowires. We have observed Pauli blockade in the double dot regime and achieved a spin qubit with long spin relaxation time ~0.6 ms at zero field. The characterization of spin states and spin lifetime presented here underscores heterostructure nanowires as a promising platform for coherent spintronics that do not suffer from fluctuating polarizations of nuclear spins and take advantage of electrical control.

**Methods Summary**

The Ge/Si core/shell heterostructure nanowires were synthesized by a nanocluster-catalyzed methodology described previously[15] and predominantly exhibit <110> growth direction. The Ge core diameter can be controlled from 10 to 20 nm by the choice of gold catalyst, and the epitaxial Si shell thickness from 2 to 5 nm. Devices are fabricated on a degenerately doped Si substrate with 600 nm thermal oxide that was grounded during measurements. Electrical contact to the nanowire was established via source/drain electrodes patterned by electron-beam lithography from a thermally evaporated film of Pd (30 nm). The entire chip was then covered by a 10 nm thin layer of $HfO_2$ (dielectric constant ~23) using atomic layer deposition. $HfO_2$ was deposited at 200°C in 100 cycles of 1 s water vapour pulse, 5 s $N_2$ purge, 3 s precursor and 5 s $N_2$ purge. Tetrakis (dimethylamino) hafnium [Hf(N(CH$_3$)$_2$)$_4$] was used as precursor. Electron-beam lithography was used to define the top gates, followed by thermal evaporation of Al (50 nm). The device was measured in a $^3$He refrigerator with a base temperature of 280 mK, and a magnetic field was applied along the axis of the nanowire within an accuracy of 30°. A Tektronix AWG520 pulse generator was used to apply fast voltages to gates L and R via low-temperature bias-Tees, resulting in a rise time of 3 nanoseconds.

**Acknowledgement**

We thank Hugh Churchill, Jim Medford and Emmanuel Rashba for technical help and discussions. We acknowledge support from the DARPA/QuEST program. Correspondence and requests for materials should be addressed to C.M.L. (cml@cmliris.harvard.edu), and C.M.M. (marcus@harvard.edu).

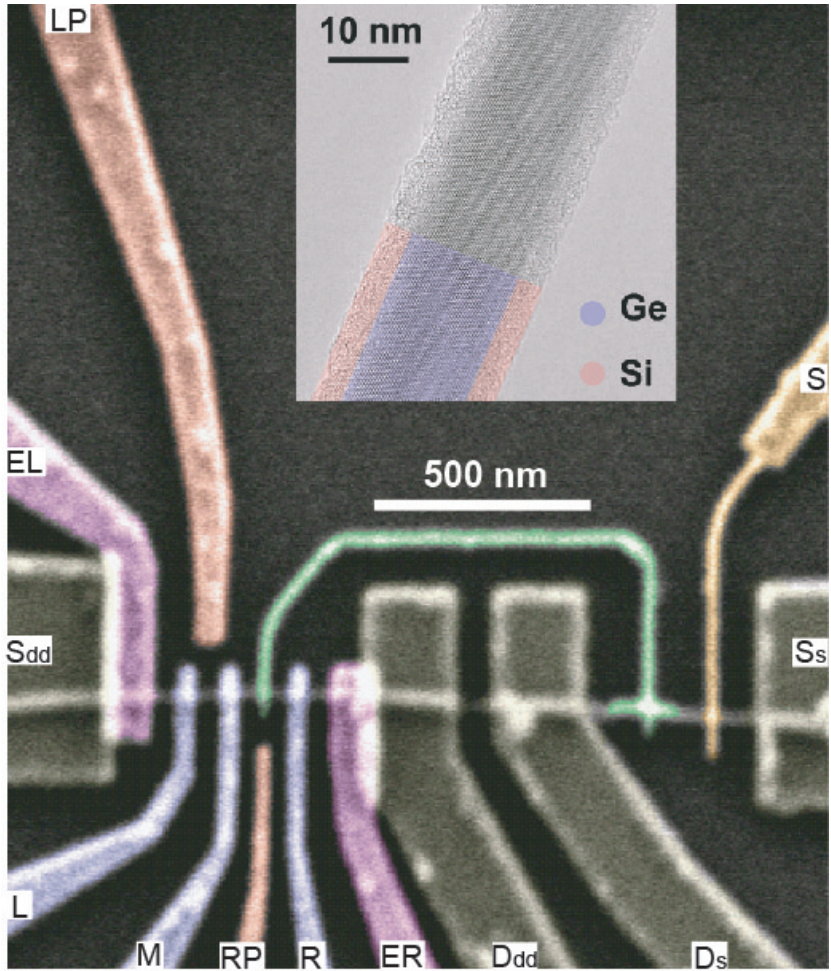

**Figure 1 | Spin qubit device based on a Ge/Si heterostructure nanowire.** Scanning electron micrograph of a Ge/Si nanowire (horizontal) contacted by four Pd contacts (grey) and covered by a $HfO_2$ gate dielectric layer. Top gates (blue) induce a double quantum dot on the left device. Plunger gates (pink) change the chemical potential of each dot independently, and side gates (purple) improve electrical contact to the nanowire. A single quantum dot on the right half of the nanowire – isolated by chemical etching between $D_{dd}$ and $D_s$ – is capacitively coupled to a floating gate (green) and tuning gate (yellow), and senses the charge state of the double dot. Inset, transmission electron microscope image of a typical nanowire with a single-crystal Ge core and an epitaxial Si shell.

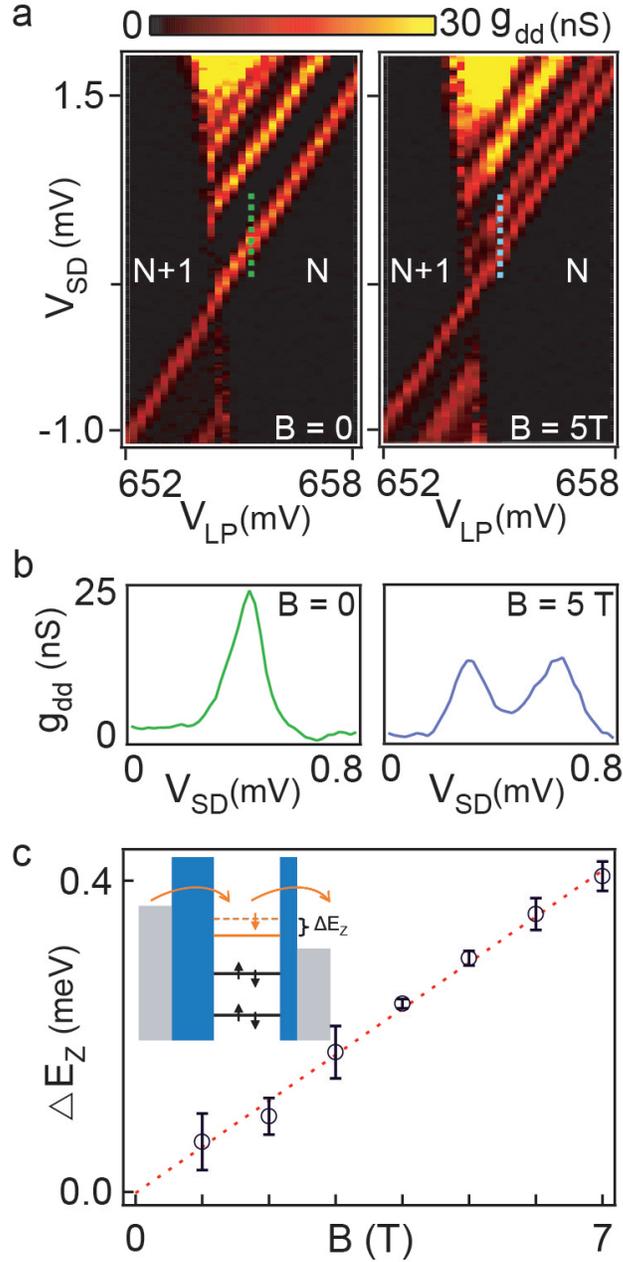

**Figure 2 | Zeeman splitting of confined holes in a single quantum dot.** a, Differential conductance $g_{dd}$ as a function of source drain bias $V_{SD}$ and gate voltage $V_{LP}$. Bright features with $V_{SD} > 0$ correspond to discrete quantum states of N+1 holes (N=even) in a single dot formed between gates L and M. b, Slices of $g_{dd}$ along dashed lines in a ($V_{LP} \sim 655$ mV) reveals Zeeman splitting of the N+1 ground state for a magnetic field of B=5 T. c, Zeeman splitting $\Delta E_Z$ versus $B$ and a linear fit (dashed line) yield a g-factor of 1.02±0.05.

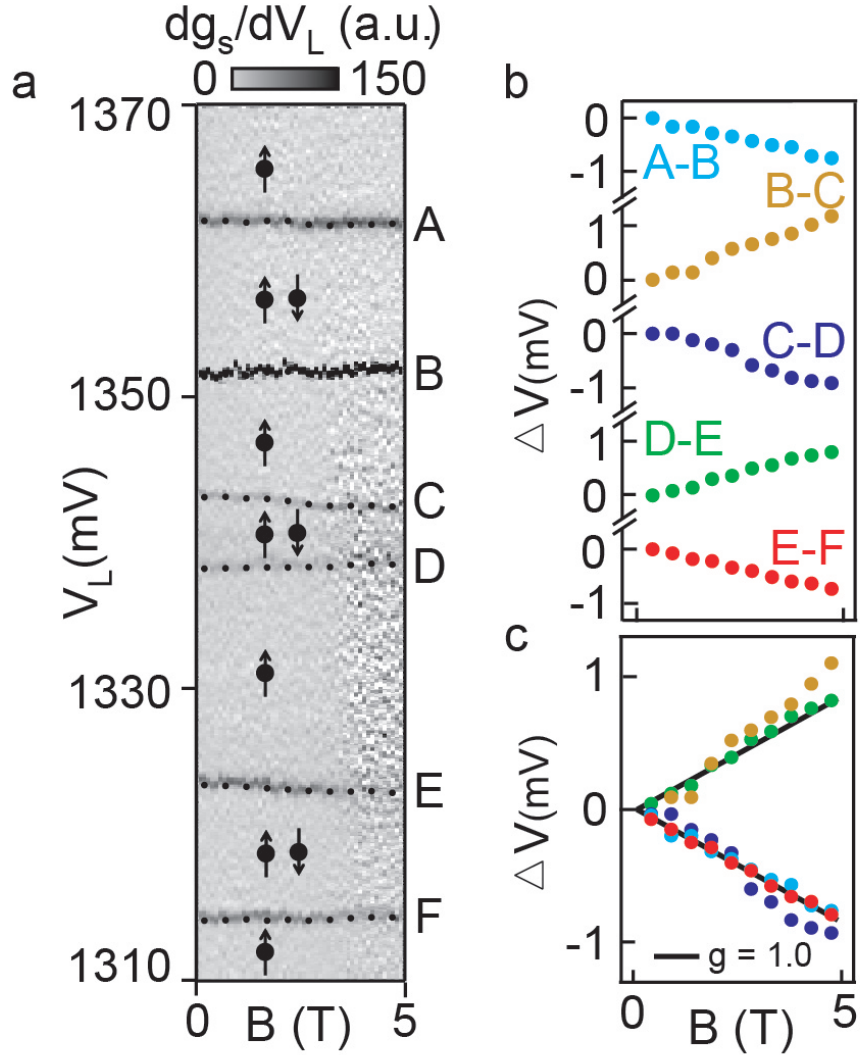

**Figure 3 | Hole spin doublets in a Ge/Si double dot.** a, Differential conductance $dg_s/dV_L$ through the sensor dot versus $B$ in the absence of current through the double quantum dot (source/drain bias = 0). Peaks in $dg_s/dV_L$ versus $V_L$ indicate groundstate transitions when holes are removed from the left dot. b, $B$ dependence of reduced Coulomb spacings, $\Delta V(B) = [V_N(B) - V_{N+1}(B)] - [V_N(B \approx 0) - V_{N+1}(B \approx 0)]$, where $V_N$ are the peak ordinates (emphasized by black dotted line in a). c, data of b plotted with guide lines $g = 1.0$ assuming a gate coupling efficiency $\alpha=0.37$ extracted from single dot device in Fig. 2.

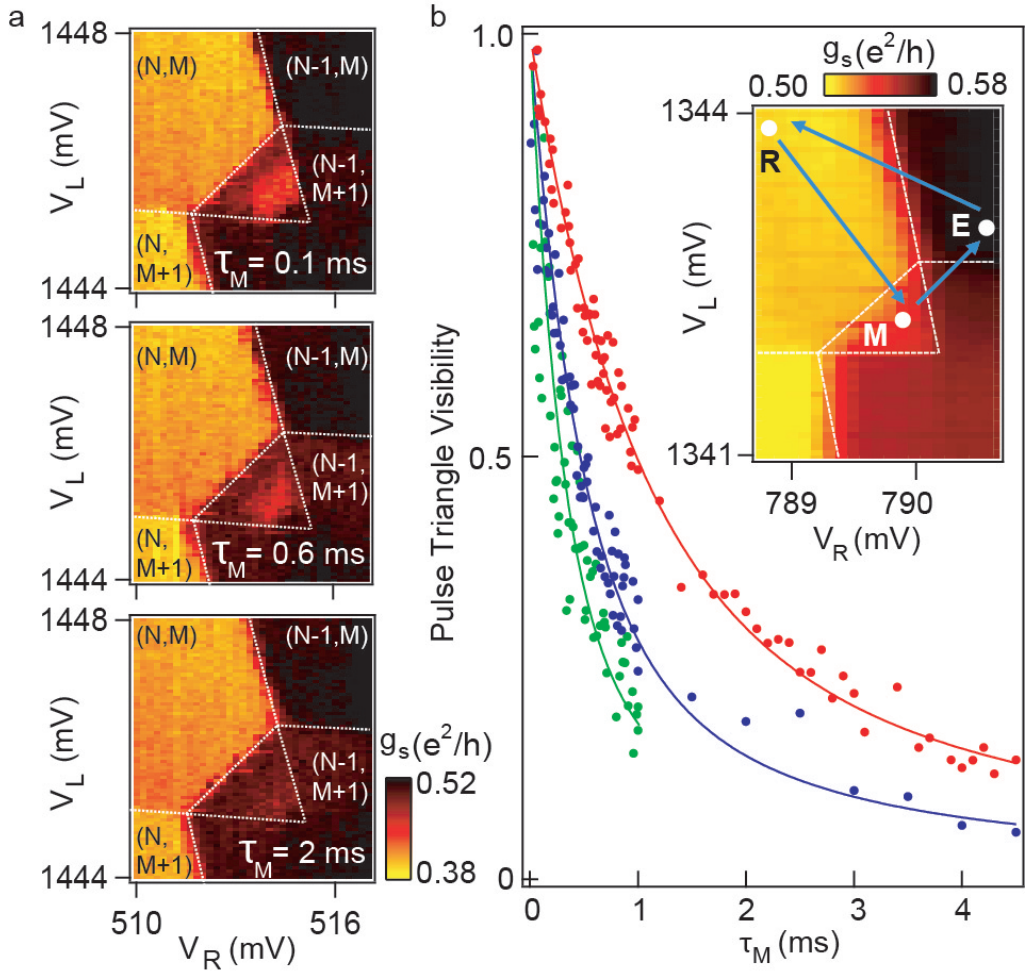

**Figure 4 | Pulsed gate measurements of spin relaxation times. a,** Sensor conductance $g_S$ near a spin-blocked charge transition between left and right dot. Spin-to-charge conversion results in pulse triangles that fade away with increasing measurement time $\tau_M$. Here N and M indicate an odd number of holes in the left and right dot (denoted as (1,1) in the main text). **b,** Visibility $I(\tau_M)$ measured at the center of the pulse triangle versus $\tau_M$ at different magnetic fields. The fitting curves (solid lines) give $T_1$=0.6, 0.3, and 0.2 ms at $B$=0 (red), 0.1 (blue), and 1 T (green) respectively. Inset, blue arrows visualize the $T_1$ pulse sequence in gate voltage space when the measurement point is held in the center of the pulse triangle.